\documentclass[conference]{IEEEtran}
\usepackage{subcaption}
\IEEEoverridecommandlockouts
\usepackage{cite}
\usepackage{amsmath,amssymb,amsfonts}
\usepackage{algorithmic}
\usepackage{graphicx}
\usepackage{textcomp}
\usepackage{pgfplots}
\usepackage{xcolor}

\usepackage[T1]{fontenc}
\def\BibTeX{{\rm B\kern-.05em{\sc i\kern-.025em b}\kern-.08em
    T\kern-.1667em\lower.7ex\hbox{E}\kern-.125emX}}
\pgfplotsset{compat=1.18}
\begin{document}
\setlength{\columnsep}{0.201in}

\title{Extracting Range-Doppler Information of Moving Targets from Wi-Fi Channel State Information\\

\thanks{Part of this work received funding from the EC Horizon Europe SNS JU projects 6G-SENSES (GA 101139282) and MultiX (GA 101192521).}
}

\author{\IEEEauthorblockN{Jessica Sanson}
\IEEEauthorblockA{\textit{Intel Deutschland GmbH} \\
Munich, Germany \\
jessica.sanson@intel.com}
\and
\IEEEauthorblockN{Rahul C. Shah}
\IEEEauthorblockA{\textit{Intel Labs} \\
Santa Clara, CA, USA \\
rahul.c.shah@intel.com}
\and
\IEEEauthorblockN{Maximilian Pinaroc}
\IEEEauthorblockA{\textit{Intel Labs} \\
Santa Clara, CA, USA \\
maximilian.c.pinaroc@intel.com}
\and
\IEEEauthorblockN{Valerio Frascolla}
\IEEEauthorblockA{\textit{Intel Deutschland GmbH} \\
Munich, Germany \\
valerio.frascolla@intel.com}
}

\maketitle

\begin{abstract}
This paper presents, for the first time, a method to extract both range and Doppler information from commercial Wi-Fi Channel State Information (CSI) using a monostatic (single transceiver) setup. Utilizing the CSI phase in Wi-Fi sensing from a Network Interface Card (NIC) not designed for full-duplex operation is challenging due to (1) Hardware asynchronization, which introduces significant phase errors, and (2) Proximity of transmit (Tx) and receive (Rx) antennas, which creates strong coupling that overwhelms the motion signal of interest. We propose a new signal processing approach that addresses both challenges via three key innovations: Time offset cancellation, Phase alignment correction, and Tx/Rx coupling mitigation. Our method achieves cm-level accuracy in range and Doppler estimation for moving targets, validated using a commercial Intel Wi-Fi AX211 NIC. Our results show successful detection and tracking of moving objects in realistic environments, establishing the feasibility of high-precision sensing using standard Wi-Fi packet communications and off-the-shelf hardware without requiring any modification or specialized full-duplex capabilities.
\end{abstract}

\begin{IEEEkeywords}
Active Sensing, Wi-Fi Sensing, ISAC, Range-Doppler Estimation.
\end{IEEEkeywords}

\section{Introduction}
Wi-Fi is ubiquitous in wireless communications \cite{b0} and while it is primarily designed for data transmission, it presents an opportunistic platform for sensing applications via Joint/Integrated Sensing and Communication (JSAC/ISAC). This dual functionality leverages existing infrastructure to provide both connectivity and environmental awareness, marking a significant advancement in wireless technology.

Active Wi-Fi sensing represents a particularly promising direction, enabling radar-like capabilities by using the same Network Interface Card (NIC) as both a transmit and a receive device. This is achieved by using one antenna as a transmitter (Tx) and another as a receiver (Rx) simultaneously, which is possible using modern Wi-Fi NICs. However, despite this ability of simultaneous transmit and receive, extracting accurate range information from commercial Wi-Fi NICs has proven challenging due to two fundamental limitations: the close proximity between Tx and Rx antennas creates overwhelming self-interference that masks the weaker reflections from objects of interest \cite{b4}, and hardware asynchronization arising from the lack of previous design considerations for full-duplex mode (radar-like) operation and relaxed requirements for communications usages. While previous work has demonstrated various sensing capabilities using Wi-Fi signals, no prior solution has successfully determined the range and Doppler of moving objects using Wi-Fi Channel State Information (CSI) in commercial devices.

In this paper we present the first demonstration of simultaneous range and Doppler estimation using CSI data from a commercial Wi-Fi NIC operating in an active sensing mode, while maintaining normal communication functions. Our approach combines delay alignment, phase offset cancellation, and Direct Current (DC) component removal techniques. Our key contribution is a new signal processing pipeline that achieves accurate range-Doppler estimation while addressing multiple technical challenges due to commercial Wi-Fi implementations:

\begin{itemize}
    \item Mitigation of self-interference between closely-spaced antennas on the same device
    \item Correction of phase and delay synchronization issues arising from the non-duplex design of the NIC
    \item Compensation for limited bandwidth constraints affecting range resolution.
\end{itemize}

 The proposed techniques are validated with real measurements from a commercially available off-the-shelf (COTS) NIC device on a commercial laptop. To validate the range and Doppler estimation values, the results are compared with those from a commercial high-resolution mmWave radar designed for short-range applications.

\section{Use Cases and Related Work}

The ability to extract range and Doppler information from commercial Wi-Fi NICs opens up numerous possibilities for innovative applications. Typically Wi-Fi sensing can detect motion parameters, such as presence/absence of movement, motion frequency etc., but cannot directly sense the location of the motion. Key parameters such as range (distance to target) and Doppler (relative velocity) can enable advanced use cases in various fields. For example, in health monitoring, Wi-Fi sensing can track respiratory and heart rates \cite{b7} but additionally, our technique could be used to directly detect breathing of two or more people without requiring techniques such as multi-source separation \cite{Multisense}. On the other hand, getting range and Doppler information can also enable intuitive control through hand movements and gesture recognition \cite{b10}. The system in \cite{Daqing-Gesture} presents gesture recognition using Wi-Fi CSI features. Similarly, \cite{applsci-11-03329} proposes WiGR, a lightweight few-shot learning framework that enables gesture recognition with low computational overhead and high cross-domain adaptability. These systems show the expanding range of use cases for Wi-Fi sensing beyond mere presence detection. This capability has a transformative impact on smart environments, particularly in Mixed-Reality (XR) applications, where motion and presence sensing can improve user experience. Activity recognition \cite{b9} benefits from these capabilities to detect and classify human actions, and presence monitoring \cite{b5} allows for detecting the presence of individuals in a room, thus enhancing security and efficiency. These applications show the potential for leveraging existing Wi-Fi infrastructure for non-intrusive, cost-effective sensing.

Methods like SpotFi \cite{b1} can extract range information but only between active Wi-Fi transceivers, not allowing to determine the distance of passive moving objects. Authors in \cite{b2} propose another approach to the range estimation of passive moving objects, i.e., using a bistatic configuration. However, this method cannot work with current COTS Wi-Fi devices since it requires one Rx antenna to be oriented towards the Tx device and capturing the direct line-of-sight reference signal, while the other Rx antenna is oriented towards the target. Some systems \cite{b3}\cite{b6} successfully extract Doppler information but cannot determine range in either monostatic or bistatic configurations. Additionally, there have been innovative indirect approaches to try to determine the location of motion (e.g.\cite{WiDir} and \cite{WiBorder}) but they are limited in their ability to extract accurate distance information.

In line with these advances, our proposed range-Doppler estimation method enables reliable gesture-level and human presence sensing through range and Doppler features, potentially empowering future  applications where Wi-Fi sensing complements vision systems for robust interaction modeling.

\section{System Model}

IEEE 802.11ac 
is a packet-based transmission scheme in which each packet comprises a preamble followed by data-carrying Orthogonal Frequency-Division Multiplexing (OFDM) symbols. The preamble is used for synchronization, channel estimation, and for identifying the length and format of the subsequent data sequence, among other functions. In particular, it contains two fields: short training symbols (STF) and long training symbols (LTF) \cite{wu2005maximum}. In this paper, we focus on the first OFDM LTF symbol in the IEEE 802.11 frame. Given that our system is a monostatic full-duplex configuration, these LTF symbols can also be used for radar-like channel estimation.

\begin{figure}[h]    \centering
    \includegraphics[width=\linewidth]{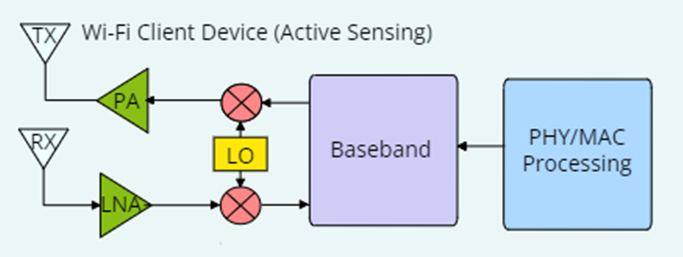}
    \caption{Wi-Fi transceiver architecture – Active sensing}
    \label{fig:wifi_transceiver_architecture}
\end{figure}

The Wi-Fi active sensing mode considered herein operates in monostatic mode, employing one Tx antenna and one Rx antenna on the same device, as illustrated in Figure~\ref{fig:wifi_transceiver_architecture}.
Active sensing introduces a radar-like operational mode that enables both range and Doppler estimation, in addition to conventional channel monitoring. Let us first consider an ideal OFDM system. The complex envelope of the accumulated $M$ transmitted Wi-Fi OFDM LTF frames is expressed as \cite{wu2005maximum}:

\begin{equation}
x(t) = \sum_{m=0}^{M-1} \sum_{n=0}^{N-1} S(m,n) e^{j2\pi n \Delta f t} \text{rect} \left( \frac{t-mT}{T} \right)
\end{equation}
where $N$ is the number of subcarriers, $M$ is the number of LTF frames, $\Delta f$ is the subcarrier spacing, $T$ is the LTF frame interval, and $S(m, n)$ represents the transmitted data or pilot on subcarrier $n$ during LTF frame $m$.

Before reaching the receiver, the OFDM LTF signal is reflected by $K$ targets. For each target $k \in K$, the signal experiences a delay $\tau_k$, which is proportional to the range of the target $r_k$, and a Doppler frequency shift $f_{D,k}$, which is proportional to its velocity $v_k$. These parameters are given by:

\begin{align}
\tau_k &= \frac{2r_k}{c}, & \text{and} \quad f_{D,k} &= \frac{2v_k f_c}{c}.
\end{align}

The received signal is expressed as:

\begin{equation}
y(t) = \sum_{k=1}^{K} x(t-\tau_k) e^{j2\pi f_{D,k} t}
\end{equation}

Putting this together, 

\begin{equation}
y(t) = \sum_{k=1}^{K} \sum_{m=0}^{M-1} \sum_{n=0}^{N-1} S(m,n) e^{j2\pi (n \Delta f + f_{D,k})(t-\tau_k)} + \tilde{\eta}(t)
\end{equation}
where $\tilde{\eta}(t)$ represents the additive white Gaussian noise (AWGN).
Hence the received OFDM symbol is:

\begin{equation}
\hat{S}(m,n) = S(m,n) \sum_{k=1}^{K} a_k e^{j2\pi T f_{D,k} m} e^{-j2\pi n \Delta f \tau_k} + \tilde{\eta}
\end{equation}

where $ \hat{S}(m,n) $ is the received OFDM symbol at subcarrier $ n $ and LTF frame $ m $, $ S(m,n) $ is the transmitted data at subcarrier $ n $ and LTF frame $ m $, $ K $ is the number of targets, $ a_k $ is the gain for target $ k $, $ f_{D,k} $ is the Doppler frequency shift for target $ k $, and $ \tau_k $ is the delay for target $ k $.

It is evident that the channel distortions are entirely encapsulated in the received modulation symbol. By comparing the transmitted and received signals (ignoring noise), we can construct the frequency-domain channel matrix:

\begin{equation}
D(m,n) = \frac{\hat{S}(m,n)}{S(m,n)} = \sum_{k=1}^{K} e^{j2\pi T f_{D,k} m} e^{-j2\pi n \Delta f \tau_k} + \tilde{\eta}
\end{equation}

Thus, the estimation of the round-trip delay and Doppler shift—and consequently, the determination of range and relative velocity of the targets—can be formulated as a spectral estimation problem. These parameters are obtained via a two-dimensional Fourier transform (2D-DFT/FFT).
\subsection{Range and Doppler Resolution}

The range resolution (ability to distinguish separate objects at different distances), $\Delta r$, is determined solely by the total bandwidth $B$ of the transmitted signal and is given by \cite{wu2005maximum}:

\begin{equation} \Delta r = \frac{c}{2B} = \frac{c}{2N \Delta f}, \end{equation}

where $B = N \Delta f$.

The Doppler resolution depends on the number of LTF frames $M$ and the time interval $T$ between frames (i.e., the frame rate), and is expressed as:

\begin{equation} \Delta v = \frac{c}{2M f_c T} = \frac{c \Delta f}{2M f_c}. \end{equation}

\subsection{Unambiguous Range and Velocity Limitations}

The maximum unambiguous range and velocity of the system are given by: \cite{sanson2020ofdm}

\begin{align}
R_{\text{max}} &= \frac{cN}{2B}, & V_{\text{max}} &= \frac{c}{2f_c T}.
\end{align}

\subsection{Radar Accuracy Estimation versus Resolution}

The measurement accuracy indicates how precisely a parameter—such as speed or range—can be determined. The accuracy of a distance measurement primarily depends on the noise level relative to the signal impulse, which is characterized by the signal-to-noise ratio (SNR). Both the noise level and the pulse edge slope (which is influenced by the bandwidth) affect measurement accuracy. For SNR $ >> 1$, the range measurement error, $\delta R$, can be approximated by \cite{skolnik2008radar}: 

\begin{equation} \delta R = \frac{c}{2B \sqrt{2 \times \text{SNR}}}, \end{equation}



Alternatively, expressing the accuracy in terms of the range resolution yields:

\begin{equation} \delta R = \frac{\Delta r}{\sqrt{2 \times \text{SNR}}}. \end{equation}

\section{Enabling Range/Doppler Estimation from Commercial NIC}


As shown in the previous section, the received OFDM signal contains both delay and Doppler information imparted by the propagation channel, however, commercial Wi-Fi systems suffer from additional practical issues. The first are time and frequency offsets introduced by the NIC hardware. Secondly, in monostatic systems with co-located Tx and Rx antennas such as in active Wi-Fi sensing, self-interference (SI) significantly limits performance. Next we discuss how we overcome these issues.


\subsection{Time-Phase Synchronization}
The starting point of our time-phase synchronization algorithm leverages the inherent Tx/Rx coupling signal of the system as a calibration reference. Since the transmit and receive antennas are co-located, the coupling path exhibits an effective delay of approximately zero. Moreover, as both antennas remain static during operation, the phase of the Tx/Rx coupling signal remains stable across frames. Given that this coupling signal is the strongest reflection received at the receiver, it serves as a reliable reference for estimating and correcting both time delays and phase offsets. This enables robust frame-to-frame alignment without requiring external calibration targets or reference signals.

We first describe how we can correct the offset problem. The initial synchronization process is based on the estimation of time and phase offsets. The information on which this process operates is the LTF sequence. Our synchronization procedure comprises of three steps: coarse time synchronization, fine time synchronization, and single-phase synchronization. For coarse time synchronization, we compute the cross-correlation between the known training-symbol sequence $S(m,n)$ and the received signal $\hat{S}(m,n)$ using
\begin{equation}
    C(l) = \sum_{n} S^*(m,n)\,\hat{S}(m,n+l),
\end{equation}
and determine the initial delay estimate as $ l_{\text{coarse}} = \arg\max_{l} \, C(l)$. Subsequently, to achieve sub-sample resolution, the signals are upsampled by a factor $U$ and a refined correlation is computed,

\begin{equation}
    C_{\text{fine}}(l) = \sum_{n} S_{\text{up}}^*(m,n)\,\hat{S}_{\text{up}}(m,n+l_{\text{coarse}}+l),
\end{equation}
yielding a fine delay estimate $  l_{\text{fine}} = \arg\max_{l} \, C_{\text{fine}}(l)$ such that the effective delay is $l_{\text{eff}} = l_{\text{coarse}} + l_{\text{fine}}$.

\begin{figure}[h]
    \centering
    \begin{subfigure}{0.49\linewidth}
        \centering
       \includegraphics[width=\linewidth]{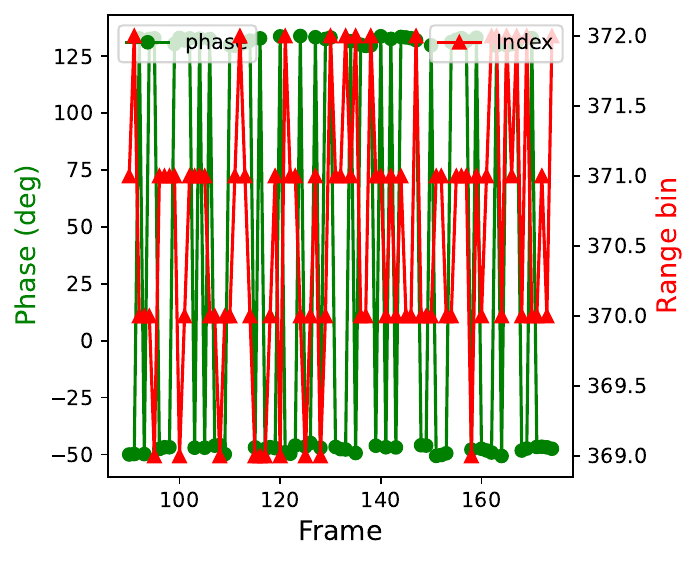}
       \caption{ }
       \label{fig:range_phasewrong}
   \end{subfigure}
   \hfill
   \begin{subfigure}{0.49\linewidth}
        \centering
       \includegraphics[width=\linewidth]{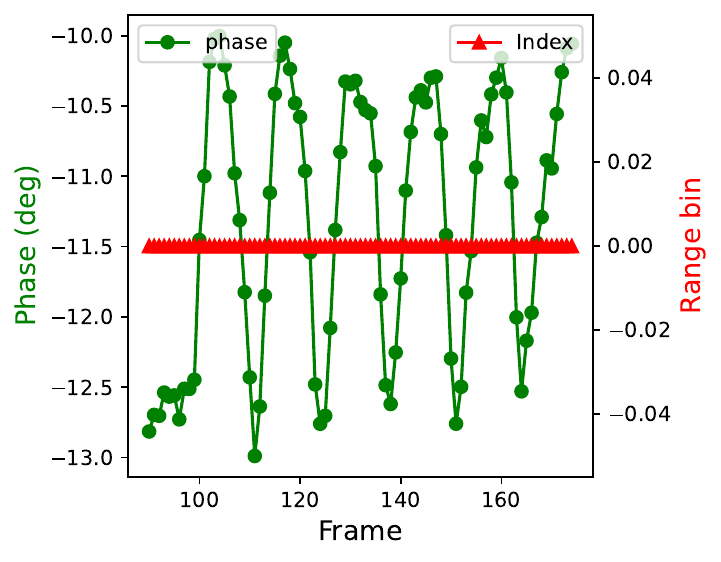}
        \caption{}
        \label{fig:range_phase_right}
    \end{subfigure}
    \caption{Before (a) and After (b) Time-Phase Synchronization}
    \label{fig:range_phase}
\end{figure}

After compensating for the time offset, the pilot symbols are removed, resulting in the extraction of the CSI. However, to leverage multiple CSI frames for Doppler estimation, we also need phase synchronization across the CSI frames. As illustrated in Figure~\ref{fig:range_phasewrong}, the phases of the CSI across frames in our system often exhibit abrupt discontinuities. The figure displays the phase evolution and corresponding range bin index versus frame number, prior to the application of time-phase synchronization. These values are extracted at the range bin corresponding to the maximum peak of the estimated cross-correlation between the known training-symbol sequence $S(m,n)$ and the received signal $\hat{S}(m,n)$. To correct these, we set the phase of the first frame as a reference and remove large phase jumps using a rounding function, which cancels abrupt changes while preserving small variations.



Let \(D(m,n)\) denote the complex CSI value at subcarrier $n$ in the $m$th frame, and let $N$ be the total number of subcarriers. The average phase of the $m$th CSI frame is computed as
\begin{equation}
\theta_m = \angle \left(\frac{1}{N} \sum_{n=1}^{N} D(m,n)\right).
\end{equation}
A reference phase $\phi_{m-1}$ is computed as the average of the corrected phases from the previous $H$ frames (with $H$ denoting the maximum history length). The phase difference between this reference phase and the current frame is then given by
$\Delta \theta_m = \phi_{m-1} - \theta_m$. To mitigate large phase jumps, the phase difference is quantized by selecting the nearest multiple of $\delta$. The correction factor is computed as
\begin{equation}
\text{fix}_m = \operatorname{round}\left(\frac{\Delta \theta_m}{\delta}\right) \cdot (\delta).
\end{equation}
Finally, the phase of the $m$th CSI frame is uniformly corrected across all subcarriers by
\begin{equation}
D(m,n) \leftarrow D(m,n) \cdot \exp\left( j\, \text{fix}_m \right).
\end{equation}
This procedure aligns the phases across CSI frames, as shown in Figure~\ref{fig:range_phase}, enabling robust Doppler estimation. Additionally, the figure illustrates the successful delay calibration, where the maximum correlation peaks are now consistently aligned at range bin zero, confirming accurate time synchronization.

\subsection{Self-Interference Cancellation}

Following time and phase synchronization, the next step is self-interference cancellation, needed in monostatic Wi-Fi sensing systems due to the fact that direct leakage from Tx to Rx can significantly distort the channel estimation. Additionally static objects generate extensive channel responses, complicating the accurate identification of moving targets. To see the effect of self-interference, we can take the example of commercial NICs implementing Wi-Fi 6 which typically have a maximum bandwidth of around 160 MHz. This results in a range resolution of approximately 0.93 m which means that the short-range Tx-Rx leakage compromises sensing of any target in the first range cell, i.e., up to 0.93 m. Hence to mitigate both self-interference and static clutter while enhancing the detection of moving targets, we propose DC component removal to cancel the zero-frequency component.

In a monostatic configuration, the coupling between the Tx and the Rx antennas appears as a zero-frequency component in the FFT/DFT results used for Doppler estimation. To address this, the DC component is removed by computing and subtracting the complex mean of the range data for each range bin across all frames. This procedure eliminates zero-Doppler power signals corresponding to static targets and Tx coupling, thereby isolating the signals from moving targets. Specifically, the modified channel matrix is given by

\[
\hat{D}(m,n) = D(m,n) - \frac{1}{M} \sum_{m=0}^{M-1} D(m,n).
\]

\section{Testbed Setup}
To evaluate the Wi-Fi sensing capabilities, we conduct two distinct tests:

\textbf{Test 1 -- Metal Tracking:} This is performed in a controlled environment to validate the accuracy of range and Doppler estimations. The setup involves a metal plate undergoing controlled motion, starting at a range of 0.6 m and ending at 0.3 m over a duration of 3.9 seconds, resulting in an approximate velocity of 0.075 m/s.

 \textbf{Test 2 -- Human Gesture:} We also explore the system's capability for gesture detection by tracking hand movements performed in front of a commercial laptop. In this scenario, a human subject moves their hand back and forth multiple times across a range of approximately 0 to 40 cm, simulating a repetitive gesture pattern. The Infineon BGT60TR13C Frequency Modulated Continuous-Wave (FMCW) radar is used as a reference for validation.

\begin{figure}[h]
    \centering
    \begin{subfigure}{0.8\linewidth}
        \centering
        \includegraphics[width=0.6\linewidth]{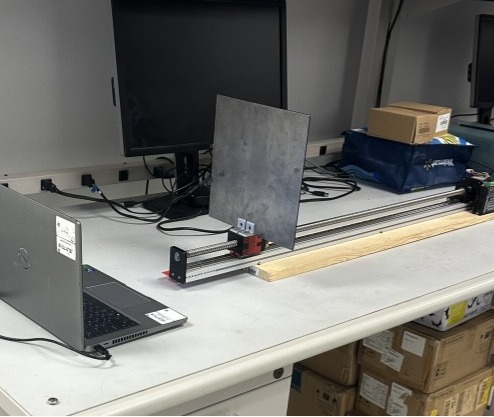}
        \caption{Controlled Tracking Setup for Test 1 -- Metal Tracking}
        \label{fig:usa_setup}
    \end{subfigure}
    \begin{subfigure}{0.48\linewidth}
        \centering
        \includegraphics[width=\linewidth]{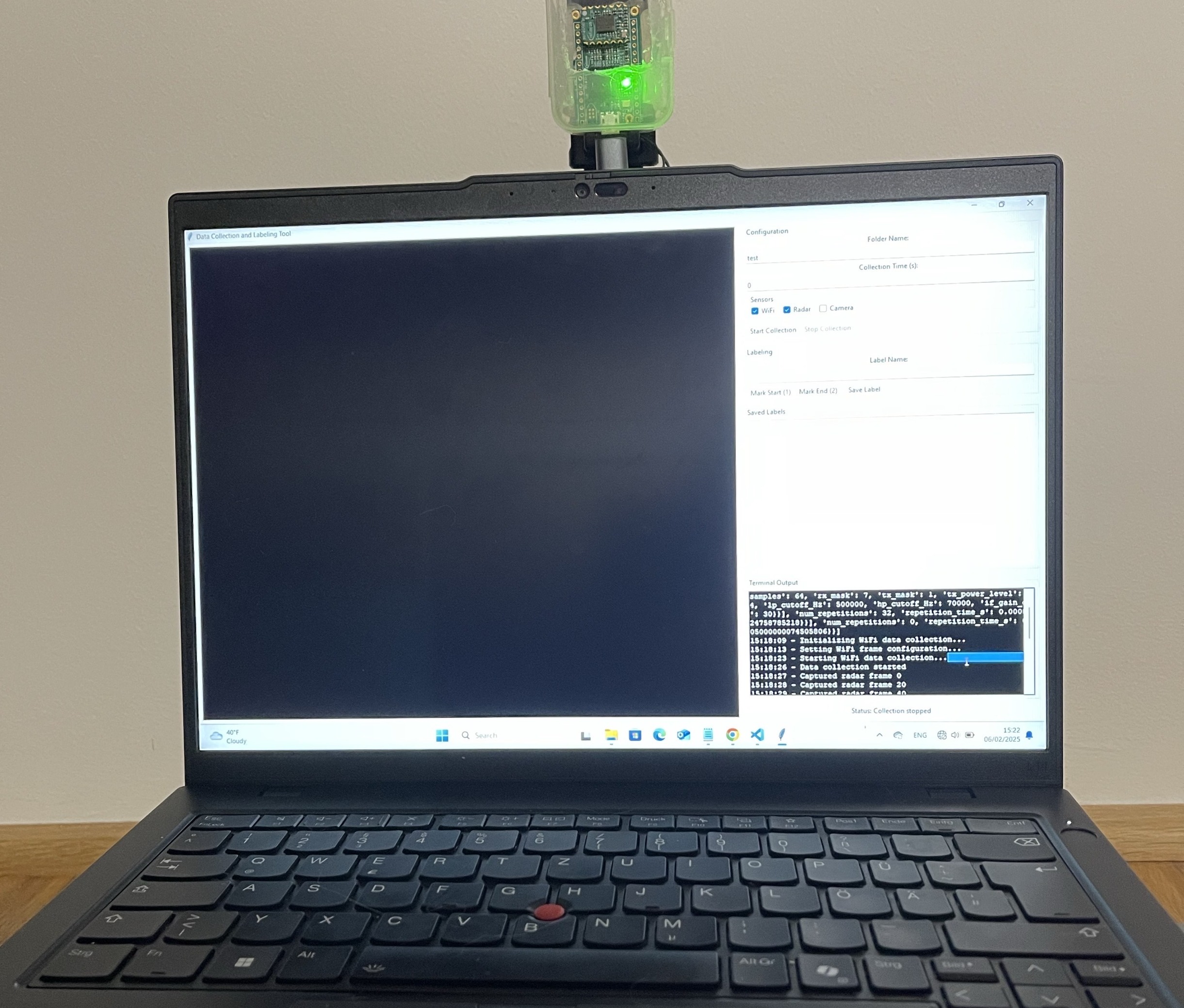}
        \caption{Device Setup for Test 2 -- Human Gesture}
        \label{fig:laptop}
    \end{subfigure}
    \hfill
    \begin{subfigure}{0.48\linewidth}
        \centering
        \includegraphics[width=\linewidth]{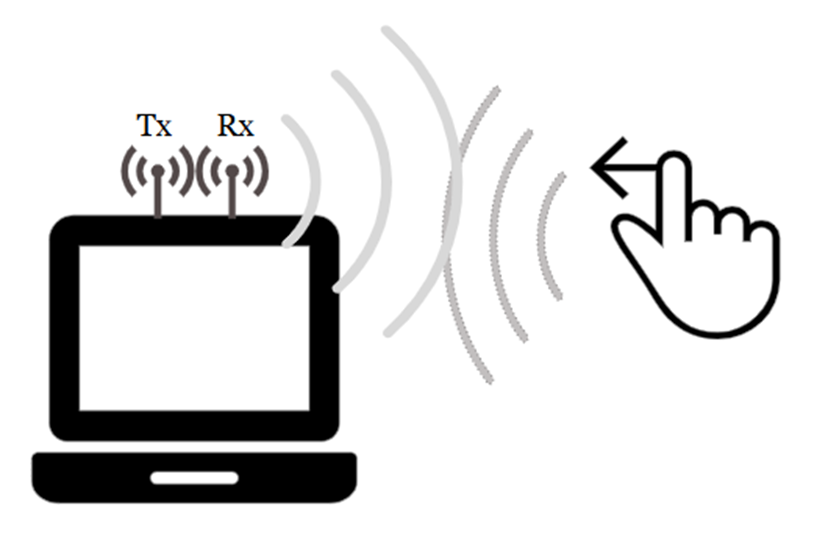}
        \caption{Hand Motion Illustration in Gesture Test}
        \label{fig:laptop2}
    \end{subfigure}
    \caption{Measurement Setup: (a) Metal Tracking (Test 1), (b) and (c) Human Gesture (Test 2)}
    \label{fig:measurement_setup}
\end{figure}

Figure~\ref{fig:measurement_setup} depicts the full experimental setup. Figure~\ref{fig:usa_setup} illustrates Test 1 (metal tracking), while Figure~\ref{fig:laptop} and Figure~\ref{fig:laptop2} illustrate Test 2, showing both the physical device setup and the gesture motion diagram, which includes the FMCW mmWave radar. The position offset from the Wi-Fi/radar system is measured and calibrated on the data. The waveform parameters for the Wi-Fi signal (used in Tests 1 and 2) and the radar (used in Test 2) are summarized in Table~\ref{tab:system_parameters}. 

LTF data is collected from one antenna at the Rx using an Intel device driver which captures CSI samples for each received packet from the card. The laptop used in both tests is a Lenovo L14 ThinkPad G5 Ultra 7 with 32GB RAM, equipped with the default Intel Wi-Fi 6E 802.11ax (AX211) Wi-Fi module. The Wi-Fi antennas are located in the bezel at the top of the display on either side of the laptop camera.
 
\begin{table}[h]
\centering
\caption{System Parameter Configurations}
\label{tab:system_parameters}
\begin{tabular}{|c|c|c|}
\hline
\textbf{Parameter} & \textbf{Wi-Fi} & \textbf{Radar} \\ \hline
Bandwidth & 160 MHz & 4 GHz \\ \hline
Carrier Frequency& 6.3 GHz & 66 GHz \\ \hline
Frame/Ramp Interval & 0.025 s & 0.0006 s \\ \hline
Range Resolution & 0.94 m & 0.03 m \\ \hline
Velocity Resolution & 0.03 m/s & 0.07 m/s \\ \hline
Max. Range & 480 m & 1 m \\ \hline
Max. Velocity & $\pm0.5$ m/s & $\pm2.24$ m/s \\ \hline
Subcarriers/Samples & 512 & 64 \\ \hline
Frames/Ramps & 32 & 64 \\ \hline
\end{tabular}
\end{table}

To the best of our knowledge, no prior method has demonstrated accurate range estimation using monostatic COTS Wi-Fi NICs without hardware modifications. Due to this absence of existing benchmarks or comparable implementations, we validate our results against a high-resolution commercial radar system. This comparison establishes a reliable ground truth and underscores the feasibility of achieving radar-like range-Doppler sensing capabilities with standard Wi-Fi hardware.

\section{Results}
\subsection{Self-Interference Cancellation}

\begin{figure}[h]
    \centering
    \begin{subfigure}{0.49\linewidth}
        \centering
        \includegraphics[width=\linewidth]{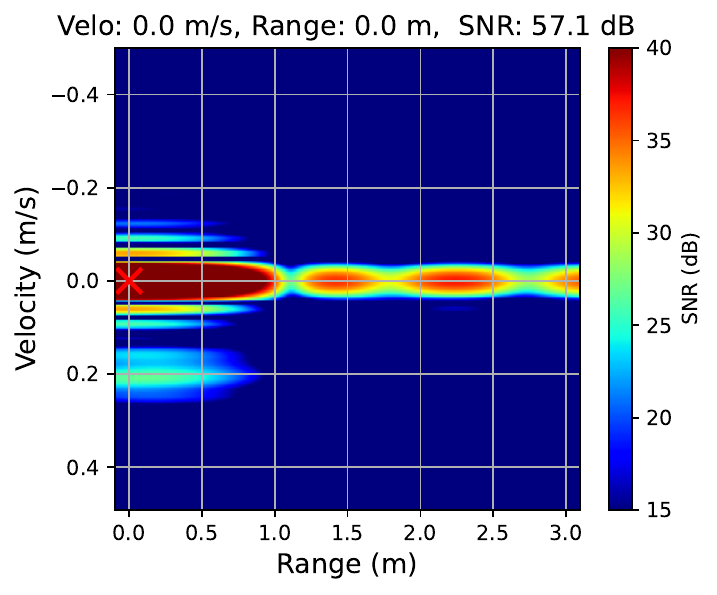}
        \caption{Without SI Cancellation}
        \label{fig:range_doppler_without_si}
    \end{subfigure}
    \hfill
    \begin{subfigure}{0.49\linewidth}
        \centering
        \includegraphics[width=\linewidth]{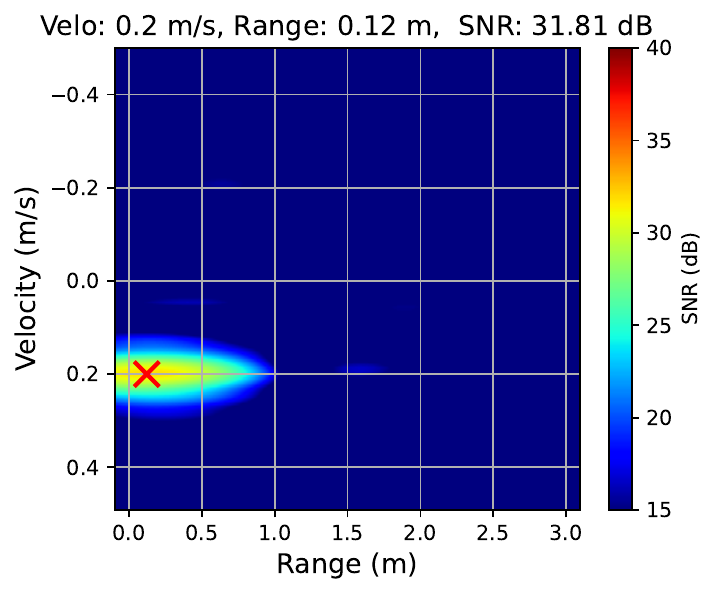}
        \caption{With SI Cancellation}
        \label{fig:range_doppler_with_si}
    \end{subfigure}
    \caption{Range-Doppler Map Estimation with/without SI Cancellation}
    \label{fig:range_doppler_si}
\end{figure}

Figure~\ref{fig:range_doppler_without_si} shows the range-Doppler map estimation for a single frame based on CSI Wi-Fi data without SI cancellation. In this case, the strong Tx/Rx coupling signal completely obscures the real target, making target detection challenging. In contrast, Figure~\ref{fig:range_doppler_with_si} presents the range-Doppler map after applying the SI cancellation technique. With the SI mitigated, the moving target becomes clearly distinguishable, allowing for accurate estimation of its range and Doppler values.

\subsection{Gesture Doppler Pattern Recognition}

\begin{figure}[h]
\centering
\includegraphics[width=0.95\linewidth]{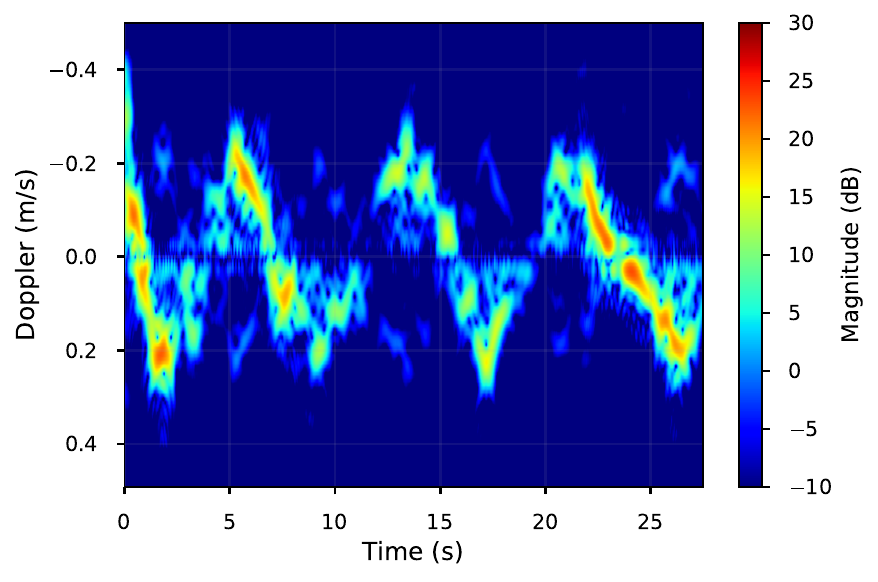}
\caption{Frequency-Time Doppler Profile of Repetitive Hand Gestures (Test 2 -- Human Gesture)}
\label{fig:gesture_velocity_heatmap}
\end{figure}

Figure~\ref{fig:gesture_velocity_heatmap} presents the frequency-time Doppler profile captured during gesture repetitions (Test 2). These plots illustrate the dynamic signature of hand gestures in the Doppler domain. Notably, the gesture introduces distinguishable positive and negative Doppler shifts corresponding to motion toward and away from the Wi-Fi antennas. The repetitive nature of the hand movement forms a periodic Doppler signature that can be easily observed and separated from noise and background clutter.

This demonstrates that Wi-Fi CSI sensing can effectively capture gesture motion patterns, revealing its potential for contactless and device-free gesture recognition. These insights lay the groundwork for more complex gesture modeling and classification tasks in future work.

\subsection{Range and Doppler Estimation}

\begin{figure}[h]
    \centering
    \includegraphics[width=0.99\linewidth]{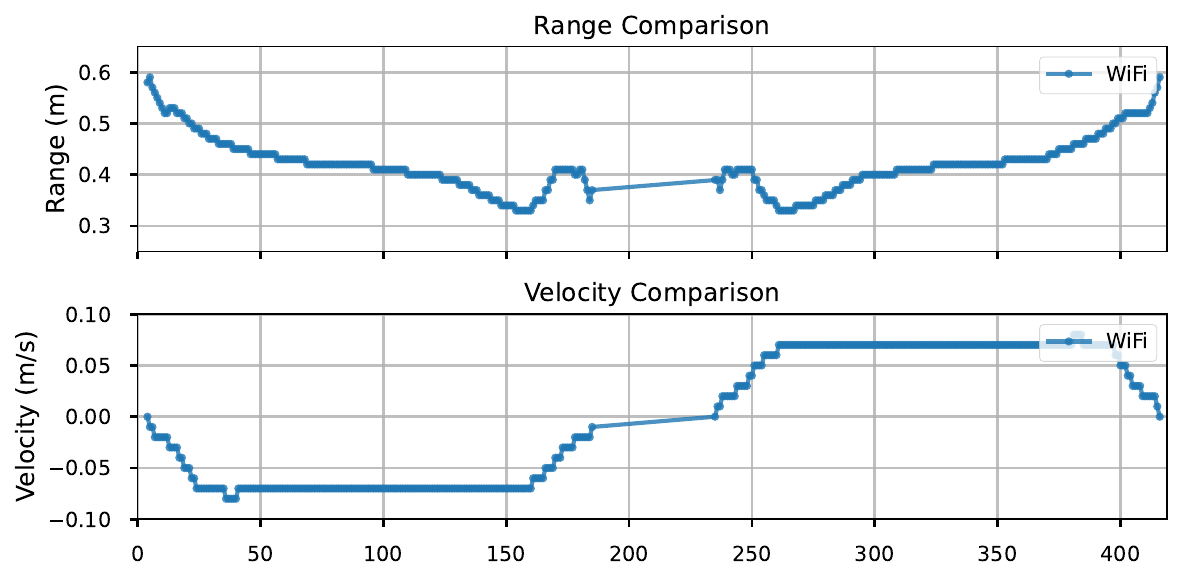}
    \caption{Range and Doppler Estimation in Test~1}
    \label{fig:range_doppler_test1}
\end{figure}

Figure~\ref{fig:range_doppler_test1} shows the corresponding range and Doppler estimation from Test~1. The Wi-Fi sensing system successfully estimates the range and Doppler information from the moving metal target while in motion. A small deviation in accuracy is noted at low velocity frames. This loss in performance is expected because the SI cancellation technique removes all static reflections, which also partially removes energy from targets at a low velocity. However, for velocities greater than the velocity resolution cell, the accuracy obtained is at the cm level for range and velocity estimation, even with a bandwidth of only 160\,MHz.

\begin{figure}[h]
    \centering
    \includegraphics[width=0.99\linewidth]{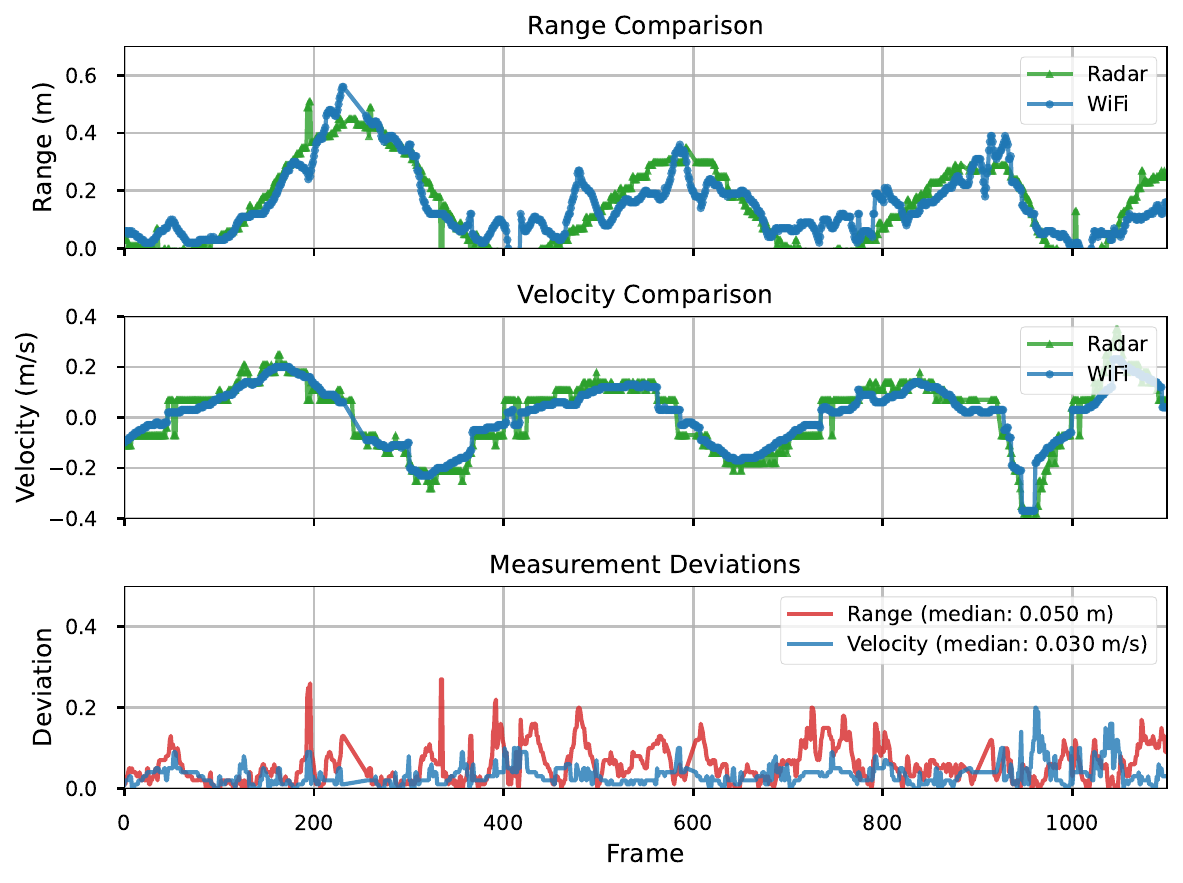}
    \caption{Range and Doppler Estimation in Test~2 -- Human Gesture}
    \label{fig:results_test2}
\end{figure}

Figure~\ref{fig:results_test2} presents the results for Test~2, where Doppler and range estimations from the Wi-Fi system are compared with those from a high-resolution radar system. In this test, the target is a human hand performing repetitive gestures, and the Wi-Fi system captures these motions via the reflected CSI signals. The comparison indicates that the range-Doppler estimation of the Wi-Fi system is accurate, with only minor deviations due to the inherent limitations of both systems. Despite the difference in resolution (0.93 m for Wi-Fi versus 0.03 m for radar), the range estimation is done with high accuracy, with median deviation errors between Wi-Fi and radar of 0.05 m and 0.03 m/s. These results clearly demonstrate the capability of the proposed technique to reliably estimate range and Doppler information using commercial Wi-Fi/devices.


\section{Conclusion}

This paper demonstrates the feasibility of Wi-Fi-based range and Doppler estimation using commercial NICs. We propose a self-interference mitigation technique to improve target detection in monostatic setups, enabling accurate estimation of moving objects even in the presence of strong Tx/Rx coupling.  Experimental results, partially obtained under the work of the MultiX and 6G-SENSES EU-funded projects \cite{6G-SENSES}, show accurate tracking of moving objects, with median errors of 0.03 m/s in velocity and 0.05 m in range, even in real-world scenarios. Our findings highlight the potential of leveraging Wi-Fi for simultaneous communication and sensing without additional hardware. This work opens new possibilities for transforming everyday Wi-Fi devices into sophisticated sensing platforms, advancing practical ISAC systems for widespread deployment.


\begin{thebibliography}{00}

\bibitem{b0} V. Frascolla, D. Cavalcanti, R. Shah, ``Wi-Fi Evolution: The Path Towards Wi-Fi 7 and Its Impact on IIoT``, in Journal of Mobile Multimedia, 19(01), 263–276, 2023.

\bibitem{b4} M. Mohammadi, Z. Mobini, D. Galappaththige, and C. Tellambura, ``A Comprehensive Survey on Full-Duplex Communication: Current Solutions, Future Trends, and Open Issues,'' IEEE Commun. Surv. Tutor., vol. 25, no. 4, pp. 1234--1267, Fourth Quarter 2023.

\bibitem{b7} Y. Gu, X. Zhang, Z. Liu, and F. Ren, ``WiFi-based real-time breathing and heart rate monitoring during sleep,'' in Proc. IEEE Global Commun. Conf. (GLOBECOM), 2019, pp. 1--6.

\bibitem{Multisense}
Y. Zeng, D. Wu, J. Xiong, J. Liu, Z. Liu and D. Zhang, ``MultiSense: Enabling Multi-person Respiration Sensing with Commodity WiFi,'' 
\emph{Proc. ACM Interact. Mob. Wearable Ubiquitous Technol.}, 2020.

\bibitem{b10} X. Yu, B. Li, and J. Chen, ``WiFi-Enabled Gesture Recognition Using Attention-enhanced DenseNet,'' in Proc. IEEE/CIC Int. Conf. Commun. China (ICCC), Hangzhou, China, 2024, pp. 1692--1697.


\bibitem{Daqing-Gesture}
R. Gao, W. Li, Y. Xie, E. Yi, L. Wang, D. Wu and D. Zhang, ``Towards Robust Gesture Recognition by Characterizing the Sensing Quality of WiFi Signals,'' \emph{Proc. ACM Interact. Mob. Wearable Ubiquitous Technol.}, 2022.



\bibitem{applsci-11-03329}
P. Hu, C. Tang, K. Yin and X. Zhang, ``WiGR: A Practical Wi-Fi-Based Gesture Recognition System with a Lightweight Few-Shot Network,'' \emph{Applied Sciences}, vol. 11, no. 8, p. 3329, 2021.

\bibitem{b9} J. Chen, K. Yang, X. Zheng, S. Dong, L. Liu, and H. Ma, ``WiMix: A Lightweight Multimodal Human Activity Recognition System based on WiFi and Vision,'' in Proc. IEEE Int. Conf. Mobile Ad Hoc Smart Syst. (MASS), Toronto, ON, Canada, 2023, pp. 406--414.

\bibitem{b5} D. Wu et al., ``WiTraj: Robust Indoor Motion Tracking With WiFi Signals,'' IEEE Trans. Mobile Comput., vol. 22, no. 5, pp. 3062--3078, May 2023.




 
\bibitem{b1} M. Kotaru, K. Joshi, D. Bharadia, and S. Katti, ``SpotFi: Decimeter level localization using WiFi,'' in Proc. ACM SIGCOMM Conf., pp. 269--282, August 2015.


\bibitem{b2} F. Colone, D. Pastina, P. Falcone, and P. Lombardo, ``WiFi-Based Passive ISAR for High-Resolution Cross-Range Profiling of Moving Targets,'' IEEE Trans. Geosci. Remote Sens., vol. 52, no. 6, pp. 3486--3501, June 2014.


\bibitem{b3} Y. Zeng, D. Wu, J. Xiong, E. Yi, R. Gao, and D. Zhang, ``FarSense: Pushing the Range Limit of WiFi-based Respiration Sensing with CSI Ratio of Two Antennas,'' Proc. ACM Interact. Mob. Wearable Ubiquitous Technol., vol. 3, no. 3, pp. 1--26, September 2019.
\bibitem{b6} F. Shi, W. Li, C. Tang, P. Brennan, and K. Chetty, ``Doppler Sensing Using WiFi Round-Trip Channel State Information,'' in Proc. IEEE Wireless Commun. Netw. Conf. (WCNC), Glasgow, UK, 2023, pp. 1--6.



\bibitem{WiDir}D. Wu, D. Zhang, C. Xu, Y. Wangand H. Wang, ``WiDir: walking direction estimation using wireless signals,''\emph{Proc. ACM Interact. Mob. Wearable Ubiquitous Technol.}, 2016.

\bibitem{WiBorder}
S. Li et al., ``WiBorder: Precise Wi-Fi based Boundary Sensing via Through-wall Discrimination,'' \emph{Proc. ACM Interact. Mob. Wearable Ubiquitous Technol.}, 2020.


\bibitem{wu2005maximum}
Y.-C. Wu, K.-W. Yip, T. S. Ng, and E. Serpedin, ``Maximum-likelihood symbol synchronization for IEEE 802.11a WLANs in unknown frequency-selective fading channels,'' \emph{IEEE Transactions on Wireless Communications}, vol. 4, pp. 2751-2763, 2005.




\bibitem{sanson2020ofdm}
J. Sanson, P. Tome, D. Castanheira, A. Gameiro, and P. Monteiro, ``High-Resolution Delay-Doppler Estimation Using Received Communication Signals for OFDM Radar-Communication System,'' \emph{IEEE Transactions on Vehicular Technology}, vol. 69, no. 10, pp. 11693--11707, Oct 2020.
\bibitem{skolnik2008radar}
M. I. Skolnik, \emph{Radar Handbook}, 3rd ed. New York, NY, USA: McGraw-Hill, 2008.

\bibitem{6G-SENSES}
J. Guti{\'e}rrez et al., ``Seamless Integration of Efficient 6G Wireless Technologies for Communication and Sensing Enabling Ecosystems,'' in \emph{Artificial Intelligence Applications and Innovations}. AIAI 2024 IFIP WG 12.5 International Workshops, 2024.


\end{thebibliography}

\end{document}